\documentclass[pdflatex,sn-nature]{sn-jnl}

\usepackage{graphicx}%
\usepackage{multirow}%
\usepackage{amsmath,amssymb,amsfonts}%
\usepackage{amsthm}%
\usepackage{mathrsfs}%
\usepackage[title]{appendix}%
\usepackage{xcolor}%
\usepackage{textcomp}%
\usepackage{manyfoot}%
\usepackage{booktabs}%
\usepackage{algorithm}%
\usepackage{algorithmicx}%
\usepackage{algpseudocode}%
\usepackage{listings}%

\theoremstyle{thmstyleone}%
\theoremstyle{thmstyletwo}%

\theoremstyle{thmstylethree}%

\begin{document}

\title[Autonomous Physical Inference of Sub-galactic Mass Substructure via Kinematically-Constrained Gravitational Lensing AI(Need update to emphasize AI)]{LensAgent: A Self Evolving Agent for Autonomous Physical Inference of  Sub-galactic Structure.
}


\author*[1,2]{\fnm{Xiaotang} \sur{Feng}}\email{xiaotang.feng@stcatz.ox.ac.uk}
\equalcont{These authors contributed equally to this work.}

\author*[1]{\fnm{Zihan} \sur{Wang}}\email{zihan.wang@queens.ox.ac.uk}
\equalcont{These authors contributed equally to this work.}

\author[1]{\fnm{Zilang} \sur{Shu}}\email{zilang.shu@pmb.ox.ac.uk}

\author[3]{\fnm{Jean-Paul} \sur{Kneib}}\email{jean-paul.kneib@epfl.ch}
\author*[2]{\fnm{Philip} \sur{Torr}}\email{philip.torr@eng.ox.ac.uk}
\affil*[1]{\orgdiv{Department of Physics}, \orgname{University of Oxford}, \orgaddress{\street{Keble Road}, \city{Oxford}, \postcode{OX1 3PU}, \country{UK}}}
\affil[2]{\orgdiv{Department of Engineering Science}, \orgname{University of Oxford}, \orgaddress{\street{Parks Road}, \city{Oxford}, \postcode{OX1 3PJ}, \country{UK}}}
\affil[3]{\orgdiv{Institute of Physics, Laboratory of Astrophysics}, \orgname{École Polytechnique Fédérale de Lausanne (EPFL)}, \orgaddress{\street{Observatoire de Sauverny}, \city{Versoix}, \postcode{CH-1290}, \country{Switzerland}}}


\abstract{Probing dark matter distribution on sub-galactic scales is essential for testing the Cold Dark Matter ($\Lambda$CDM) paradigm. Strong gravitational lensing, as one of the most powerful approach by far, provides a direct, purely gravitational probe of these substructures. However, extracting cosmological constraints is severely bottlenecked by the mass-sheet degeneracy (MSD) and the unscalable nature of manual and neural-network modeling. Here, we introduce LensAgent, a pioneering training-free, large language model (LLM)-driven agentic framework for the autonomous physical inference of mass distributions. Operating as an autonomous scientific agent, LensAgent couples high-level logical reasoning with deterministic physical modeling tools, demonstarting successful reconstruction of mass distribution in SLACS Grade A strong lensing systems. This self-evolving architecture enables the robust extraction of sub-galactic substructures at scale, unlocking the cosmological potential of upcoming wide-field surveys such as the Rubin Observatory (LSST) and Euclid.}

\keywords{Artificial Intelligence, Autonomous Inference, Strong Gravitational Lensing, Subhalos}



\maketitle

\section*{Introduction}\label{sec1}

The nature of dark matter remains one of the most profound mysteries in modern physics. The standard $\Lambda$CDM model predicts a hierarchical universe populated by numerous low-mass sub-galactic halos \cite{Springel:2005nw,Springel:2008cc}; however, the persistent discrepancy between these predictions and the observed satellite population \cite{Planck:2018vyg}—the so-called "Small-Scale Crisis" \cite{Bullock:2017xww}—suggests either a fundamental misunderstanding of dark matter properties or a significant population of dark subhalos devoid of baryonic matter.
Strong gravitational lensing provides the only direct, purely gravitational probe to map these elusive mass distributions, independent of baryonic tracers \cite{Despali:2016meh}. By analyzing distortions in lensed arcs, localized potential perturbations caused by dark matter substructures can be identified \cite{2019MNRAS.483.5649S}. Although theoretically appealing, this approach needs to overcome two critical limitations. First is the fundamental Mass-Sheet Degeneracy (MSD) \cite{Schneider:2013sxa} that imaging data alone cannot uniquely constrain the absolute mass scale or the logarithmic density profile of the lens. Second is that extracting sub-galactic signals requires extreme precision in macro-model subtraction. In general, this method traditionally necessitates highly resource-intensive, expert-led iterative fitting — a paradigm unscalable for the $10^5$ strong lenses expected from next-generation wide-field surveys such as the Rubin Observatory (LSST) \cite{LSST:2008ijt}and Euclid \cite{Euclid:2021icp}.

To bypass this computational bottleneck, recent efforts have turned to utilize automated deep learning algorithms to infer lens parameters \cite{Shajib:2025bho,Zhang:2023wda}. While offering remarkable speed and efficiency, these supervised approaches are limited by exclusive dependence on massive volumes of synthetic training data. This introduces an unavoidable "simulation-to-reality gap", that out-of-distribution samples in mock datasets may cause the neural network to overfit on its training data, alongside the high computational overhead of training, makes it challenging to enforce strict physical symmetries or break degeneracies like the MSD without introducing unphysical priors.

In this study, we develope LensAgent, the first, to our knowledge, training-free multimodal large language model (LLM) agentic framework designed for the autonomous physical inference of mass distributions. Under the revolutionary development recently in Artificial Intelligence, Large language models encompasse a wide spectrum of general world and scientific knowledge \cite{openai2023gpt4}, and have been shown to be able to iterate and learn new tasks through few-shot learning and tool feedback \cite{FewShot,Viper,Zhang2025ExploringTR}. LensAgent robustly breaks the MSD by integrating spectroscopic kinematics and ensures physical self-consistency through numerical validation of the Poisson equation. In this work, we demonstrate the efficacy of LensAgent by applying it to 20 Grade A samples of SLACS, showcasing its frontier ability to reconstructing galatic mass distribution and dark matter subhalo detection.

\section*{LensAgent Architecture}

LensAgent couples an inner ReAct-style agent \cite{YaoReact}, augmented with explicit tool use \cite{schick2023toolformer}, and iterative self-feedback \cite{shinn2023reflexion,madaan2023selfrefine} to an island-based evolutionary outer search. \cite{Romera-Paredes2024,MeyersonNBGKHL24,AlphaEvolve} Therefore the system has three components: an inner agent that proposes and revises parameters, a proposal database organized into islands, and a deterministic evaluator that scores each proposal against imaging, kinematics and physical validity. One outer iteration consists of selecting an island, sampling context from that island, running one agent episode, evaluating the returned proposal, and writing accepted results back to the same database. The proposal retained at iteration $t$ alters the prompt seen at iteration $t+1$. This recursive update is the mechanism of self evolution.


\begin{figure*}
\centering
\includegraphics[width=\textwidth]{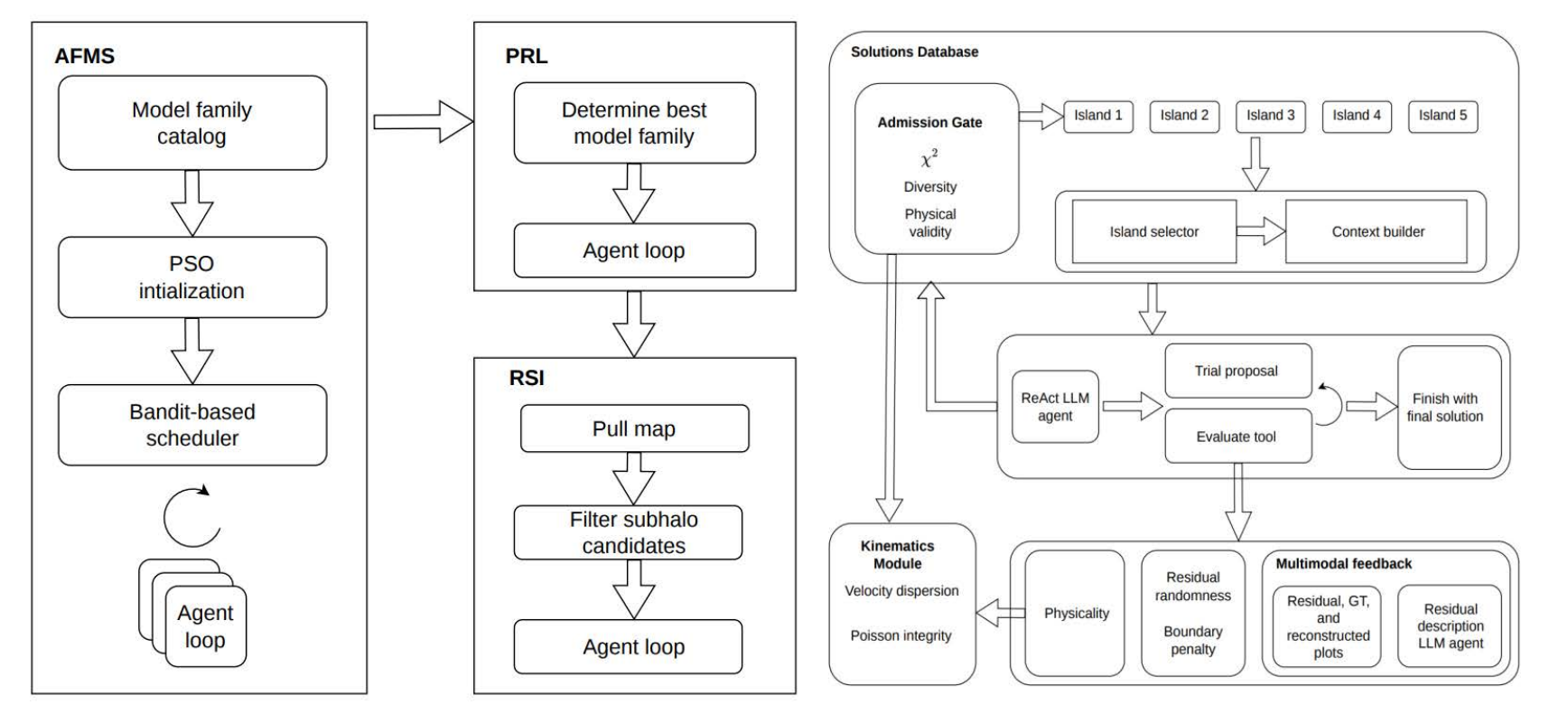}
\caption{Architecture and workflow of LensAgent. The proposal database is organized into islands and serves as the evolving search state. Context sampled from one island conditions a ReAct inner agent, which proposes three parameter sets, invokes deterministic evaluation, and returns a revised proposal. Accepted fits are written back to the same database. LensAgent is used in model family search (AFMS), precision refinement (PRL) and residual driven subhalo fitting (RSI).}
\label{fig:ai_agent_workflow}
\end{figure*}

At each iteration, the agent is prompted to submit three candidate solutions. This format improves the output diversity of the llm, by forcing it to diverge from the local optima and consider more diverse solutions. It also serves as a way to improve sampling efficiency and reduce cost by generating more solutions at each iteration.

After each \texttt{evaluate} call, the agent receives reduced image $\chi^2$, kinematic $\chi^2$, residual randomness, physicality diagnostics, and comparison panels of the data, render and residuals. To further improve the visual grounding of the model, we implement a visual analysis module that uses a seperate LLM call to provide a description of residual morphology. The agent iterates for up to $T=5$ steps, revising later proposals in response to evaluator feedback, formulated in the ReAct paradigm \cite{YaoReact}. The best candidate across the episode is returned.

We utilize LensAgent in our fitting process to perform the full workflow of model family selection, parameter optimization and subhalo detection. The process is devided into three phases: Autonomous Fitting-driven Model Selection--AFMS; Parameter Refinement Loop--PRL; Residual-based Subhalo Inference--RSI (Algorithm demonstrated in Methods \ref{alg:pipeline}).

\section*{AFMS--Autonomous Fitting-driven Model Selection}
Strong gravitational lensing is fundamentally a manifestation of the Shapiro delay and the geometric deflection of light-rays in a perturbed Friedmann-Lemaître-Robertson-Walker (FLRW) metric, where the propagation of light is governed by the two-dimensional lensing potential $\psi(\boldsymbol{\theta})$ (Detail derivation in Methods \ref{theory}). The agent navigates a comprehensive library of mass profiles model families to accurately capture the best validated $\chi^2$ of reconstruction of 3D mass profile. To accommodate the challenge of optimizing parameters and selecting the model family simultaneously, we initialize LensAgent across the candidate model families(See table \ref{tab:families} in Methods \ref{pip}) in this phase. A PSO scout finds initial solutions by running multiple PSO searches for each family and supplementing them with random initializations, then ranks the resulting fits and seeds the LensAgent databases to cold-start the run. Agent budget is then allocated across the selected families by a bandit-based scheduler using a UCB rule \cite{FINITE}. This allows a dynamic balance of exploration and exploitation, since model families with more promising validated solutions receive a higher number of invocations as the phase progresses. Within each model family, we maintain a separate island database. For enhanced efficiency and cost optimization, AFMS also uses an early-stopping mechanism at two levels. First, families whose best validated solution stops improving for a sustained patience window of 10 iterations per model family are removed from the active pool. Second, if repeated scheduler scoreboards continue to identify the same family as the leader, the model selection phase is terminated early and that family is passed to the next refinement stage. This phase autonomously identifies the optimal physical parameterization for the deflector galaxy's baryonic and dark matter distributions, preventing under-fitting while penalizing unphysical complexity.

\section*{PRL--Parameter Refinement Loop}
Locking in the highest-evidence model family from the AFMS phase, the agent executes the PRL to achieve higher numerical precision. To break the MSD, this phase explicitly links the 2D projected surface density to the 3D dynamical gravitational potential (In Methods \ref{theory}). The agent iteratively refines parameters from the model family with highest likelihood until the predicted luminosity-weighted stellar velocity dispersion perfectly anchors the spectroscopic observables. Furthermore, it strictly enforces mass-potential self-consistency by numerically validating the Poisson equation (In Methods \ref{theory}). This phase retains a patience-based early stop, now applied within the single selected model family, such that the loop halts once the best validated entry in the carried-over database ceases to improve meaningfully. Both the above phase and this phase aims to reconstruct real 3D mass profile with validated physics properties.

\section*{RSI--Residual-based Subhalo Inference}
Furthermore, We load the single best solution from the above stage. Utilizing the normalized residual from the fully converged model, the agent introduces localized NFW or Gaussian perturbations to identify mass concentrations that cannot be accounted for by the smooth profile. It isolate subhalo candidates, converting the converged residuals into pull-maps in general. Significant peaks detected in this map are used to perturb the selected model family with NFW subhalos. After selecting the most likely subhalo coordinate set, we perform PSO runs with the perturbed parameter set and use LensAgent to fit the subhalo parameters. If no significant residual candidates are found, this stage terminates immediately. Otherwise, LensAgent is rerun on the observed system with the perturbed seeds, again with a patience-based early stop on the best validated subhalo solution. Notably, we observe that it is common for the agent to over-optimize the solution, leading to unphysical massive subhalo proposals. To address this issue, proposals that violate the hard subhalo mass cap of $10^{10} M_{sun}$ are rejected before admission, and the agent only iterates on the retained population.

\section*{Results}
In this work we have presented LensAgent, an automated strong gravitational lens modeling pipeline that integrates multi-component mass model selection, lens light and source reconstruction, kinematic cross-validation, and dark matter substructure detection into a unified framework built upon \texttt{lenstronomy}. Applied to 20 Grade-A SLACS systems using archival HST/ACS F814W imaging and SDSS spectroscopy, the pipeline produces physically self-consistent mass models with minimal manual intervention.

The reduced $\chi^2$ values for the best-fit models range from 0.994 (J1103$-$5322) to 1.150 (J1250$+$0523), confirming that the automated model selection procedure identifies parameterizations of appropriate complexity for each system. The preferred configurations span a broad range: the majority of systems are well described by single elliptical power-law (EPL) profiles with external shear, consistent with the established near-isothermality of massive early-type galaxies at SLACS redshifts~\cite{Koopmans:2006iu}, while four systems  favor composite Hernquist$+$NFW decompositions and four others (J0959$+$4416,J1204$+$0358, J1205$+$4910, J1636$+$4707) require dual EPL parameterizations. More specialized configurations, including multipole perturbations (J1451$+$0239) and Gaussian source-plane clumps (J1531$-$0105), further illustrate the pipeline's adaptability.This suggest the physical detail of the galaxies for future studies.
The predicted stellar velocity dispersions agree with the independently measured SDSS values for all 20 systems within the $1\sigma$ observational uncertainties. This concordance between two fundamentally independent probes provides robust validation that the pipeline recovers mass models that are not merely good fits in the image plane but are dynamically meaningful.

We also report five promising subhalo detections. The masses lies in the regime where cold dark matter and alternative models such as warm or self-interacting dark matter make divergent predictions for subhalo abundance~\cite{Lovell:2013ola,2025arXiv251218959W}. The $\sim 5.6 \times 10^9\,M_\odot$ perturber near the Einstein radius of J1029$-$0420 is consistent with a massive satellite or dwarf galaxy-scale object expected in CDM~\cite{2015MNRAS.447.3189X}.
 \begin{table}[h]
\centering
\begin{tabular}{lcccl}
\toprule
SDSS id & Model selected & Reduced $\chi^2 $&$\sigma$ predicted(km/s)&$\sigma$ Observed(km/s)\\
\midrule
J0037+0942 & Dual EPL+Shear& 0.999 & 282&$279\pm 14$\\
J0044+0113 &SIE+Shear & 1.000&265.1&$266\pm 13$ \\
 J0405+0455& EPL+Shear& 1.040 & 161.8&$160\pm 8$ \\
 J0912+0029&EPL+Convergence&1.001&328.1&$326\pm 16$\\
 J0959+4416&Dual EPL+Shear&0.999&232.7&$244.0\pm 14$\\
 J1029-0420& EPL+SIS&0.999&210.1&$210\pm 11$\\
 J1103-5322&Hernquist+NFW+SHEAR&0.994&189.1&$196\pm 12$\\
 J1112+0826&PEMD+Shear&1.035&328.2&$320\pm 20$\\
 J1142+1001&EPL+Shear+SIS&1.018&224.4&$221\pm 22$\\
 J1153+4612&Hernquist+NFW+Shear&0.999&222.9&$226\pm 15$\\
 J1204+0358&EPL+EPL+Shear&0.999&250.4&$267\pm 17$\\
 J1205+4910&EPL+EPL+Shear&1.125&280.9&$281\pm 14$\\
 J1218+0830&EPL+Shear+SIS&1.073&219.2&$219\pm 11$\\
 J1250+0523&EPL+Shear+SIS&1.150&258.4&$252\pm 14$\\
 J1250-0135&EPL+Shear+SIS&0.999&255.4&$246\pm 12$\\
 J1451+0239&EPL+Shear+Multipole&0.999&214.4&$223\pm 14$\\
 J1531-0105&EPL+Shear+Gaussian Clumps&1.060&272.7&$279\pm 14$\\
 J1627-0053&Hernquist+NFW+Shear&0.999&294.2&$290\pm 15$\\
 J1636+4707&EPL+EPL+Shear&0.999&230.9&$231\pm 15$\\
 J2341+0000&Hernquist+NFW+Shear&0.999&207.8&$207\pm 13$\\
\bottomrule
\end{tabular}
\caption{Table for all strong lensing samples fitted, randomly from SLACS Survey's Grade A catalog.The corresponding SDSS id, model combination of best fitted, velocity dispersion predicted and velocity dispersion observed. }
\label{tab:results}
\end{table}

\begin{table}[h]
\centering
\begin{tabular}{lcccl}
\toprule
SDSS id & Mass($M_{sun}$) & center(x,y) \\
\midrule
 J0037+0942&1.47e6&2.1782,0.9901\\
 J0912+0029&5.6e6&-2.1777,1.7818\\
 J1029-0420&5.76e9&-0.5455,0.6700\\
 J1103-5322&5.95e7&-0.2413,0.1276\\
 J1451+0239&4.88e8&0.9850,-0.5890\\
\bottomrule
\end{tabular}
\caption{Potential subhalo detected by LensAgent. }
\label{tab:results}
\end{table}
We also demonstrates the fitted image against observed and the residual map.\newpage
\begin{figure}[t]
    \centering
    \includegraphics[width=0.75\linewidth]{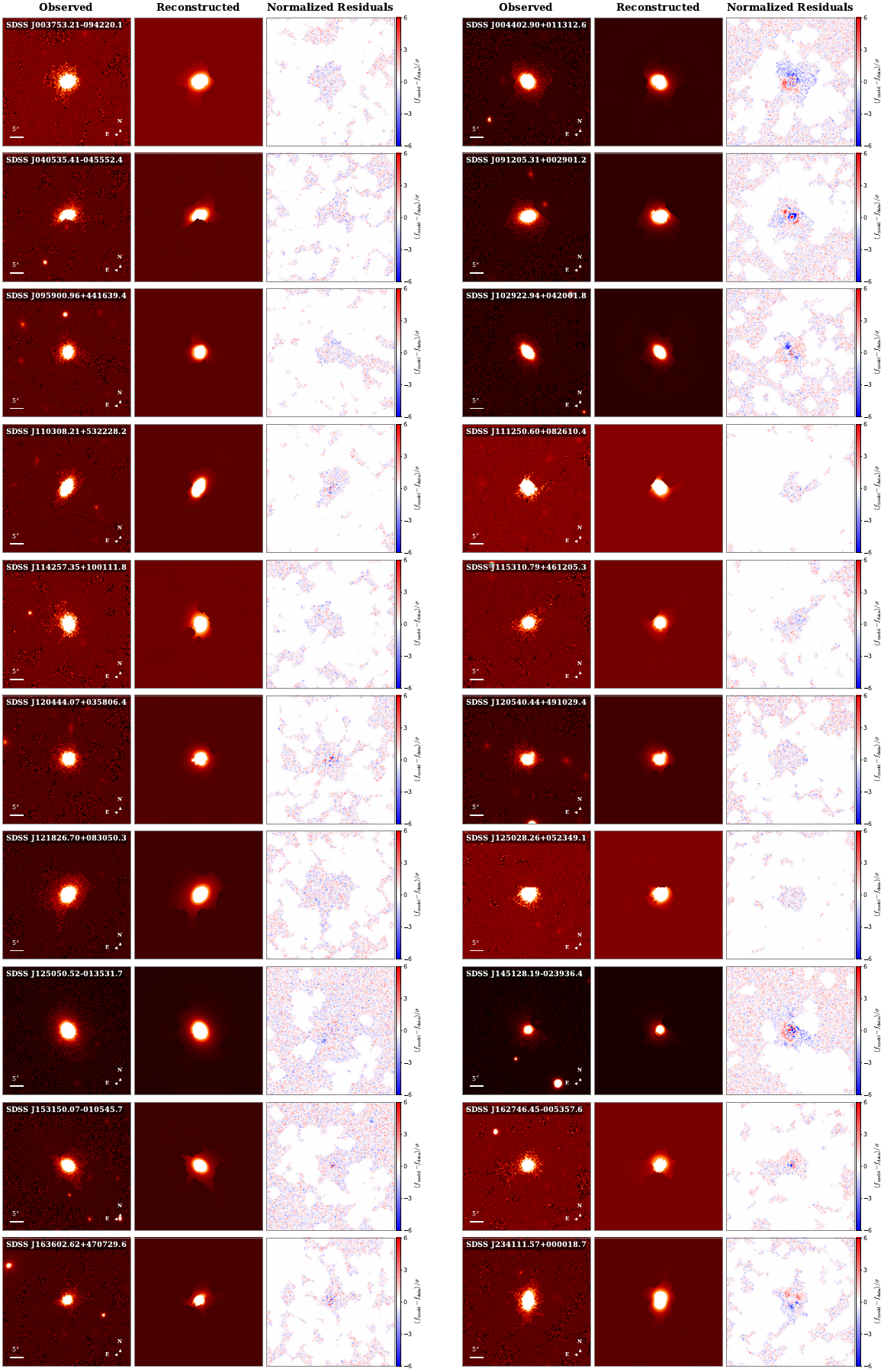}
    \caption{The whole 20 samples fitted by LensAgent. The right panels are the observed image along with SDSS id by HST. The middle panels are the reconstructed ones by lensAgent. The left panels are the normalized residual of observed subtract reconstructed. }
\end{figure}
\newpage
\section*{Discussion }
\label{sec:discussion_conclusions}

Looking ahead, Euclid and LSST are expected to discover of order $10^5$ galaxy--galaxy strong lenses~\cite{Euclid:2021icp,Collett:2015roa}, rendering manual modeling impractical. Automated pipelines such as LensAgent will be essential for extracting the full scientific return from these datasets. Several avenues for future development are envisioned: incorporating spatially resolved integral-field spectroscopy \cite{Padovani:2023dxc} to tighten constraints on radial mass profile slopes and luminous--dark matter decompositions; extending the substructure search with formal Bayesian model comparison to place statistically rigorous constraints on the subhalo mass function; and coupling the pipeline with machine-learning-based lens finders to create an end-to-end system progressing from survey images to mass models with minimal human oversight.

In summary, LensAgent autonomously models strong gravitational lens systems to a standard comparable with dedicated manual analyses, recovering accurate mass distributions validated by independent kinematic measurements while simultaneously probing the small-scale structure of dark matter halos. As large-sample strong lensing science begins in earnest, tools of this kind will be indispensable for transforming the unprecedented volume of incoming data into precise constraints on the nature of dark matter and the mass structure of galaxies.

\bibliography{sn-bibliography}

\section*{Methods}
\subsection*{Data}
\label{Data}
The lensed sample catalog is selected from SLACS survey.We have selected 20 samples that are marked as grade A, which are confirmed to be strong lensing samples.We then obtain the image of these samples observed by HST,particularly, we are using the observation in F814W. For the velocity dispersion data, we use the result of kinematic analysis from SDSS. \newline
We then preprocess the HST imaging data to isolate the lens systems for detailed morphological and mass modeling. To accurately model the noise properties of the imaging data, we perform a two-dimensional background estimation using the photutils package\cite{2023software}. First, we generate a comprehensive source mask to exclude the primary deflector, lensed arcs, and any neighboring interlopers. To prevent the extended, low-surface-brightness wings of these sources from biasing the background calculation, we expand the initial mask using a binary dilation algorithm with three iterations\cite{2020NatMe..17..261V}. We then map the local background across the field using a 2D mesh grid. Within each local box, the background level is evaluated using a median estimator, coupled with a $3 \times 3$ median filter to smooth the resulting large-scale background map. Furthermore, we apply an iterative sigma-clipping algorithm to robustly reject residual outliers, such as unmasked faint sources or cosmic rays. The resulting root-mean-square (RMS) background map provides a reliable characterization of the $1\sigma$ pixel-to-pixel noise variation, which is subsequently incorporated into the modeling pipeline to compute the $\chi^2$ likelihood of the lens models.\newline
To accurately account for the blurring effects of the telescope optics and detector response, we explicitly characterize the Point Spread Function (PSF) for both the high-resolution imaging and the spectroscopic data. For the HST/ACS F814W imaging, we generate a synthetic PSF at the detector position of each lens system using the TinyTim software package \cite{2011SPIE.8127E..0JK}, which models the full optical path of the Hubble Space Telescope including the obscuration pattern, optical aberrations, charge diffusion in the CCD, and the wavelength-dependent diffraction structure of the ACS Wide Field Channel. The resulting oversampled PSF image is rebinned to the native ACS pixel scale (0.05'' pixel), center-cropped, flux-normalized, and provided to the lenstronomy modeling framework \cite{Birrer:2018xgm} as a pixelated convolution kernel. During the likelihood evaluation, all proposed high-resolution surface brightness models are convolved with this kernel before comparison with the observed data. To account for any residual mismatch between the synthetic TinyTim PSF and the true instrumental PSF—arising,we additionally correct the PSF iteratively during the modeling process following the procedure described in \cite{Birrer:2018xgm}, whereby perturbative corrections to the PSF are reconstructed from the residuals of the lens model fit. Furthermore, to forward-model the kinematic observations obtained with the SDSS spectrograph, we utilize an analytical Moffat profile to represent the atmospheric seeing conditions of the ground-based spectroscopic data. This Moffat PSF—defined by its Full Width at Half Maximum (FWHM) and $\beta$ parameter—is applied to convolve the intrinsic Multi-Gaussian Expansion surface brightness models, ensuring that the spatial smearing of the light is accurately accounted for when calculating the luminosity-weighted line-of-sight velocity dispersion within the spectroscopic aperture.
\subsection*{Theoretical Framework of Strong Gravitational Lensing}
\label{theory}

Strong gravitational lensing is fundamentally a manifestation of the Shapiro delay and the geometric deflection of light-rays in a perturbed Friedmann-Lemaître-Robertson-Walker (FLRW) metric. The propagation of light from a distant source to an observer can be described by the Fermat potential $\tau(\boldsymbol{\theta}, \boldsymbol{\beta})$, representing the excess travel time relative to an unperturbed path:
\begin{equation}
    \tau(\boldsymbol{\theta}, \boldsymbol{\beta}) = \frac{1+z_l}{c} \frac{D_l D_s}{D_{ls}} \left[ \frac{1}{2}(\boldsymbol{\theta} - \boldsymbol{\beta})^2 - \psi(\boldsymbol{\theta}) \right]
\end{equation}
where $z_l$ is the lens redshift, and $D_l, D_s, D_{ls}$ are the angular diameter distances to the lens, the source, and between the lens and source, respectively. According to Fermat's principle, stationary points of this potential ($\nabla_{\boldsymbol{\theta}} \tau = 0$) correspond to the observed image positions, yielding the standard lens equation:
\begin{equation}
    \boldsymbol{\beta} = \boldsymbol{\theta} - \nabla \psi(\boldsymbol{\theta}) = \boldsymbol{\theta} - \boldsymbol{\alpha}(\boldsymbol{\theta})
\end{equation}
where $\boldsymbol{\alpha}(\boldsymbol{\theta})$ is the scaled deflection angle.

The local distortion of the lensed image is characterized by the Jacobian of the lens mapping, $\mathbf{A} = \partial \boldsymbol{\beta} / \partial \boldsymbol{\theta}$. This matrix can be decomposed into the convergence $\kappa$, representing isotropic focus, and the complex shear $\gamma = \gamma_1 + i\gamma_2$, representing anisotropic stretching:
\begin{equation}
    \mathbf{A} = 
    \begin{pmatrix} 
    1 - \kappa - \gamma_1 & -\gamma_2 \\ 
    -\gamma_2 & 1 - \kappa + \gamma_1 
    \end{pmatrix}
\end{equation}
The convergence $\kappa(\boldsymbol{\theta})$ is the dimensionless projected surface mass density, related to the lensing potential $\psi$ via the two-dimensional Poisson equation: $\nabla^2 \psi = 2\kappa$. The local magnification of a lensed image is given by the inverse of the determinant of the Jacobian: $\mu = [\text{det} \mathbf{A}]^{-1} = [(1-\kappa)^2 - |\gamma|^2]^{-1}$. In strong lensing, the locus of points where $\text{det} \mathbf{A} = 0$ defines the critical curves in the image plane and the caustics in the source plane, where magnification formally diverges.

For early-type galaxies, the total mass distribution is accurately described by the Elliptical Power Law (EPL) profile, often referred to as the power-law ellipsoid. The convergence of this profile is given by:
\begin{equation}
    \kappa(x, y) = \frac{3-\gamma}{2} \left[ \frac{\theta_E}{\sqrt{q x^2 + y^2 / q}} \right]^{\gamma-1}
\end{equation}
where $\theta_E$ is the Einstein radius, $q$ is the minor-to-major axis ratio, and $\gamma$ is the 3D logarithmic density slope ($\rho \propto r^{-\gamma}$). This profile generalizes the Singular Isothermal Ellipsoid (SIE, where $\gamma=2$) and allows for a flexible radial mass distribution. The integration of multipole perturbations ($a_m$) further allows the model to capture non-elliptical azimuthal deviations, such as disky or boxy structures, which are critical for eliminating systematic residuals in high-resolution data.

We choose the elliptical S\'ersic function to model the
deflector light profile. The S\'ersic function is parameterized as
\begin{equation}
    I(\theta_1, \theta_2) = I_e \, \exp\!\left[ -k \left\{ \left(
    \frac{\sqrt{\theta_1^2 + \theta_2^2 / q_L^2}}{\theta_{\mathrm{eff}}}
    \right)^{1/n_{\mathrm{Sersic}}} - 1 \right\} \right].
\end{equation}
Here $I_e$ is the amplitude, $k$ is a constant that normalizes
$\theta_{\mathrm{eff}}$ so that it is the half-light radius, $q_L$ is the axis
ratio, and $n_{\mathrm{Sersic}}$ is the S\'ersic index. The coordinates
$(\theta_1, \theta_2)$ are obtained by rotationally transforming the on-sky
coordinates $(\theta_x, \theta_y)$ relative to the lens centre
$(\theta_{x,0},\, \theta_{y,0})$ using the position angle $\phi_L$:
\begin{align}
    \theta_1 &= \phantom{-}(\theta_x - \theta_{x,0})\cos\phi_L
                + (\theta_y - \theta_{y,0})\sin\phi_L, \\
    \theta_2 &= -(\theta_x - \theta_{x,0})\sin\phi_L
                + (\theta_y - \theta_{y,0})\cos\phi_L,
\end{align}
so that $\theta_1$ and $\theta_2$ are aligned with the major and minor axes of
the elliptical isophotes, respectively. The isophotes are concentric ellipses
with constant axis ratio $q_L$ and position angle $\phi_L$. 
For the case that a single S\'ersic function that can not  model the lens light well, we introduce a multiple S\'ersic function with same centroid to achieve optimal fit.

Detection of sub-galactic mass clumps (subhalos) is treated as a perturbation problem. The total potential is decomposed into a smooth macro-model and a set of localized perturbations:
\begin{equation}
    \psi_{tot}(\boldsymbol{\theta}) = \psi_{smooth}(\boldsymbol{\theta}) + \sum_{i} \psi_{sub,i}(\boldsymbol{\theta} - \boldsymbol{\theta}_{i})
\end{equation}
Our pipeline identifies these perturbations through two automated stages.

Initial subhalo candidates are identified in the "Pull Map"—a normalized residual map calculated as $P = (D - M) / \sigma$, where $D$ is the data, $M$ is the smooth model, and $\sigma$ is the total noise (background and Poisson). Significant dipoles in the Pull Map ($> 5\sigma$) signal the presence of mass concentrations that cannot be accounted for by a smooth EPL profile.

For each identified candidate, the pipeline injects a localized mass profile such as a Singular Isothermal Sphere or a Gaussian Kappa clump) and re-optimizes.

\subsection*{Kinematic Modeling and Lensing}
\label{subsec:kinematics}

To validate the mass models recovered from the imaging analysis and to break the mass-sheet degeneracy inherent in lensing-only constraints, we predict the stellar line-of-sight velocity dispersion for each best-fit lens model and compare these predictions against the independently measured spectroscopic values. Our kinematic modeling framework is built upon the module of \texttt{lenstronomy} \cite{Birrer:2018xgm}, which implements the unified lensing and kinematic formalism developed by \cite{Shajib:2019crn}and the Jeans anisotropic modeling framework of \cite{2016ARA&A..54..597C}.

The fundamental principle underlying our kinematic analysis is that the same mass distribution responsible for the strong-lensing deflection also governs the stellar dynamics of the deflector galaxy. While lensing observables probe the projected mass enclosed within the Einstein radius, stellar kinematics are sensitive to the three-dimensional gravitational potential within the spectroscopic aperture. This complementarity enables a powerful cross-check: given a parameterized mass model constrained by the lensing data, one can predict the luminosity-weighted line-of-sight velocity dispersion $\sigma_{\rm los}$ and compare it to the observed value.

The computation of $\sigma_{\rm los}$ requires three ingredients: a three-dimensional mass density profile $\rho(r)$, a three-dimensional luminosity density profile $l(r)$, and a prescription for the stellar orbital anisotropy $\beta(r)$. Following \cite{Shajib:2019crn}, we obtain the three-dimensional profiles by deprojecting the two-dimensional surface density and surface brightness distributions under the assumption of spherical symmetry. The deprojection is made analytically tractable by decomposing each profile into a sum of Gaussian components via the Multi-Gaussian Expansion (MGE) technique. For a two-dimensional Gaussian component with amplitude $\Sigma_{0j}$, standard deviation $\sigma_j$, and projected axis ratio $q_j$, the Abel inversion yields a three-dimensional Gaussian of the form
\begin{equation}
\label{eq:rho_gaussian}
\rho_j(r) = \frac{\Sigma_{0j}}{\sqrt{2\pi}\,\sigma_j\,q_j} \exp\!\left(-\frac{r^2}{2\sigma_j^2}\right),
\end{equation}
and the three-dimensional enclosed mass takes the closed-form expression
\begin{equation}
\label{eq:mass_gaussian}
M_j(r) = \frac{2\pi\,\sigma_j^2\,\Sigma_{0j}}{q_j}\left[\mathrm{erf}\!\left(\frac{r}{\sqrt{2}\,\sigma_j}\right) - \sqrt{\frac{2}{\pi}}\,\frac{r}{\sigma_j}\exp\!\left(-\frac{r^2}{2\sigma_j^2}\right)\right],
\end{equation}
where $\mathrm{erf}(x)$ is the error function. The total mass and luminosity profiles are recovered by summing over all Gaussian components, exploiting the linearity of the Jeans equations. Under the assumption of spherical symmetry, the velocity dispersion is obtained by solving the spherical Jeans equation,
\begin{equation}
\label{eq:jeans_spherical}
\frac{\mathrm{d}\!\left(l\,\overline{v_r^2}\right)}{\mathrm{d}r} + \frac{2\beta\,l\,\overline{v_r^2}}{r} = -l\,\frac{\mathrm{d}\Phi}{\mathrm{d}r},
\end{equation}
where $\Phi$ is the gravitational potential satisfying $\mathrm{d}\Phi/\mathrm{d}r = GM(r)/r^2$, and $\beta(r) \equiv 1 - \overline{v_\theta^2}/\overline{v_r^2}$ is the velocity anisotropy parameter. The luminosity-weighted, line-of-sight second velocity moment at projected position $(x,y)$ is then given by
\begin{equation}
\label{eq:vlos_integral}
\overline{v_{\rm los}^2}(x,y) = \frac{2G}{I(x,y)}\int_{\sqrt{x^2+y^2}}^{\infty} \mathcal{K}_\beta\!\left(\frac{r}{\sqrt{x^2+y^2}}\right)\,l(r)\,M(r)\,\frac{\mathrm{d}r}{r},
\end{equation}
following \cite{2005MNRAS.362...95M}, where $I(x,y)$ is the projected surface brightness and the kernel $\mathcal{K}_\beta(\upsilon)$ encodes the anisotropy prescription. For the isotropic case ($\beta = 0$), the kernel simplifies to $\mathcal{K}_\beta(\upsilon) = \sqrt{1 - 1/\upsilon^2}$. For the Osipkov--Merritt parameterization $\beta(r) = r^2/(r^2 + r_{\rm ani}^2)$, where $r_{\rm ani}$ is the anisotropy radius, the kernel takes the more complex form given by equation~(3.18) of \cite{Shajib:2019crn}. Because each component of both the mass and light profiles is Gaussian, the enclosed mass $M(r)$ and the luminosity density $l(r)$ are expressed entirely in terms of error functions and elementary operations, eliminating the need for nested numerical integration within the Jeans equation solver.

We demonstrates the full kinematic prediction pipeline through the following sequence of operations.

First, the class is initialized with the lens and source redshifts ($z_{\rm d}$, $z_{\rm s}$), the lens model specification, the spectroscopic aperture geometry, the seeing conditions, and the choice of anisotropy model. The angular diameter distances $D_{\rm d}$, $D_{\rm s}$, and $D_{\rm ds}$ are computed from the assumed cosmology via the \texttt{LensCosmo} class and packaged into a cosmological keyword dictionary that is passed to the \texttt{Galkin} solver.

Second, the projected half-light radius $R_{\rm eff}$ of the deflector light distribution is computed numerically from the lens light model via the \texttt{LightProfileAnalysis} module, unless a pre-computed value is supplied. Similarly, the circularized Einstein radius $\theta_{\rm E}$ and, when required, the local logarithmic slope $\gamma$ of the mass profile at $\theta_{\rm E}$ are evaluated from the lens model via the \texttt{LensProfileAnalysis} module.

Third, the lens mass profiles are translated into a form compatible with the \texttt{Galkin} kinematic solver. When the MGE mode is enabled, the composite convergence profile is evaluated along a logarithmically spaced radial grid spanning $10^{-4}$ to $10^{2}$ times the Einstein radius. The resulting one-dimensional radial profile is then decomposed into a set of Gaussian components ($\sim\!20$ by default) using the \texttt{mge\_1d} routine, which implements the integral transform method of \cite{Shajib:2019crn}. This method introduces an integral transform with a Gaussian kernel,
\begin{equation}
\label{eq:gaussian_transform}
F(x) = \int_0^{\infty} \frac{f(\sigma)}{\sqrt{2\pi}\,\sigma}\exp\!\left(-\frac{x^2}{2\sigma^2}\right)\mathrm{d}\sigma,
\end{equation}
whose inverse can be computed efficiently via a modified Euler algorithm for inverse Laplace transforms. The resulting amplitudes and standard deviations of the Gaussian components are passed to the \texttt{Galkin} solver as a \texttt{MULTI\_GAUSSIAN} mass profile. Alternatively, when the mass profile admits a direct three-dimensional deprojection (e.g., a power-law or NFW profile), the lens model parameters are passed through to \texttt{Galkin} without MGE decomposition.

Fourth, the deflector light profile is similarly prepared for kinematic input. Three options are available in the pipeline: (i)~direct use of the \texttt{lenstronomy} light profile types that possess analytic three-dimensional deprojections; (ii)~a Hernquist approximation, in which the actual light distribution is replaced by a \cite{1990ApJ...356..359H} profile with scale radius $R_{\rm s} = 0.551\,R_{\rm eff}$, matched to reproduce the observed half-light radius; or (iii)~an MGE decomposition of the radial light profile, analogous to the mass profile treatment. In all cases, the ellipticity of the light profile is set to zero for the kinematic computation, as the Jeans solver operates under the assumption of spherical symmetry. The center coordinates of the light distribution are inherited from the lens light model to ensure spatial consistency.

Fifth, a \texttt{Galkin} instance is created for each spectroscopic observation, encapsulating the mass profile, light profile, anisotropy model, aperture geometry, seeing conditions, and numerical integration settings. The \texttt{Galkin} solver then computes the luminosity-weighted, aperture-averaged line-of-sight velocity dispersion by Monte Carlo integration: a large number ($N_{\rm sampling} = 1000$ by default) of stellar phase-space positions are drawn within the spectroscopic aperture, convolved with the seeing point-spread function, and the resulting line-of-sight velocity moments are averaged with luminosity weighting. For spatially resolved (IFU) kinematics, a two-dimensional velocity dispersion map can be computed using either a grid-based convolution scheme or a shell-based decomposition via the \texttt{GalkinShells} class.

Finally, the predicted velocity dispersion is corrected for any external convergence $\kappa_{\rm ext}$ not included in the lens model via the mass-sheet transformation,
\begin{equation}
\label{eq:kappa_ext_correction}
\sigma_{v,\rm corrected} = \sigma_v\,\sqrt{1 - \kappa_{\rm ext}},
\end{equation}
which accounts for the fact that an unmodeled external mass sheet rescales the inferred mass normalization and hence the predicted velocity dispersion.

For the present analysis, we adopt the following kinematic modeling configuration. We assume the Osipkov Merritt anisotropy model with the anisotropy radius treated as a free parameter, allowing us to marginalize over the unknown orbital structure of the deflector galaxy. The mass profile for kinematic modeling is constructed from the subset of lens model components that describe the main deflector, excluding external shear and any substructure perturbers; this selection is controlled by a boolean mask  applied to the full lens model list. The light profile is taken directly from the best-fit lens light model with ellipticity set to zero. We employ the MGE decomposition with 20 Gaussian components for both the mass and light profiles to ensure sub percent accuracy in the radial profile reconstruction over the range $0.1\,R_{\rm eff} \leq R \leq 10\,R_{\rm eff}$. The spectroscopic aperture is configured to match the SDSS fiber diameter of $3^{\prime\prime}$, and the seeing is set to the median SDSS spectroscopic PSF full width at half maximum. Each velocity dispersion prediction is computed with 1000 spectral rendering samples to ensure numerical convergence to better than 1 per cent.
As a final validation, the pipeline computes the numerical Laplacian of the converged potential $\psi$ on a high-resolution grid. We verify that $\nabla^2 \psi \approx 2\kappa$ across the entire image plane. Any model exhibiting a Root Mean Square Error or significant regions of negative convergence is discarded as unphysical, ensuring that numerical artifacts are not mistaken for dark matter detections.

\subsection*{Pipeline process}
\label{pip}
Here we ouline the fitting process of our pipeline.

We first extract the lensing sample in a 120 times 120 pixel cutout from the whole image by HST. We then choose stars that also in the image but not target to construct PSF of the observation. We then applied the SExtractor to estimate the background rms value for each sample. We then masked out the nearby galaxies and stars that were not part of the lensing system.

We run a particle swamp optimization across all combinations of models available, the deatiled list of them is \ref{tab:families}.
\begin{table}[h]
\centering
\begin{tabular}{@{}llp{5cm}@{}}
\toprule
\textbf{Family} & \textbf{Mass Components} & \textbf{Physical Motivation} \\
\midrule
Standard EPL          & EPL + SHEAR                  & Baseline power-law lens \\
EPL + Multipoles      & EPL + SHEAR + MULTIPOLE      & Azimuthal complexity \\
PEMD + Shear          & PEMD + SHEAR                 & Alternative ellipticity implementation \\
SIE + Shear          & SIE + SHEAR                 & Cored isothermal model \\
MGE Mass Model        & MULTI\_GAUSSIAN + SHEAR      & Kinematic compatibility \\
EPL + Convergence     & EPL + SHEAR + CONVERGENCE    & Mass-sheet / line-of-sight \\
Stars + DM (NFW)      & HERNQUIST + NFW + SHEAR      & Baryon--DM decomposition \\
Group Environment     & EPL + SHEAR + SIS            & Satellite perturbers \\
Substructure          & EPL + SHEAR + GAUSSIAN\_KAPPA& Dark matter clumps \\
Merger / Dual Centre  & EPL + EPL + SHEAR            & Complex central mass \\
\bottomrule
\end{tabular}
\caption{Model families explored by the automated scouting pipeline.}
\label{tab:families}
\end{table}
Then we calculate the fitting quality for each model's PSO runs, and initialize our solutions database for the agentic fitting phase described below.

\subsection*{Inner agent}

At the core of the framework is a tool-using LLM agent \cite{YaoReact} that interacts with the lens modeling environment through two tools, following the broader paradigm of tool-augmented language models \cite{schick2023toolformer}. The agent receives the observed image, PSF, noise model, two context proposals sampled from the selected island, and the observed stellar velocity dispersion $\sigma_v^{\mathrm{obs}}$. Because the prompt combines visual inputs with structured numerical context, the setup utilizes the multimodal capabilities of modern LLMs \cite{openai2023gpt4,liu2023visual,MMREACT}. It then reasons over parameter updates and invokes one of the following tools:

\begin{itemize}
    \item \textbf{\texttt{evaluate}} submits three parameter vectors $\{\boldsymbol{\theta}^{(1)}, \boldsymbol{\theta}^{(2)}, \boldsymbol{\theta}^{(3)}\}$. Each is rendered and scored independently.
    \item \textbf{\texttt{finish}} submits three final candidates and terminates the episode, returning the best.
\end{itemize}

At each iteration, the agent is prompted to submit three candidate solutions. This format improves the output diversity of the llm, by forcing it to diverge from the local optima and consider more diverse solutions. It also serves as a way to improve sampling efficiency and reduce cost by generating more solutions at each iteration.

After each \texttt{evaluate} call, the agent receives reduced image $\chi^2$, kinematic $\chi^2$, residual randomness, physicality diagnostics, and comparison panels of the data, render and residuals. To further improve the visual grounding of the model, we implement a visual analysis module that uses a seperate LLM call to provide a description of residual morphology. The agent iterates for up to $T=5$ steps, revising later proposals in response to evaluator feedback, formulated in the ReAct paradigm \cite{YaoReact} which is a self-reflection and self-refinement loops \cite{shinn2023reflexion,madaan2023selfrefine}. The best candidate across the episode is returned. We use \textit{Gemini-3.1-pro-preview} for the main agent, and \textit{Gemini-3.1-flash-lite-preview} for the residuals description module. We sample all LLM models with the following parameters: $\text{Temperature}=1.0$, $\text{Top\_}p=0.95$ and $\text{Reasoning\_effort}=\text{"high"}$

\subsection*{Proposal database and evolutionary search}

The inner agent is embedded in an evolutionary outer loop that combines language-model proposal generation with quality-diversity archival search \cite{MappingElites,Pugh2016QualityDA}. The proposal database is split into $I=5$ islands. Each island stores accepted proposals, their scalar evaluation results, quality scores, diversity scores and behavior vectors. In the present implementation, the database is the memory. To improve solution diversity, the islands' entries are independently stored without crossover.

At each outer iteration $t$, one island is selected from the set of non-empty islands $\mathcal{N}_t$ with a mild rank-based quality bias. Let $q_i^\star(t)$ be the best quality currently stored in island $i$, and let $r_i(t)\in\{1,\dots,|\mathcal{N}_t|\}$ denote the rank of island $i$ after sorting the non-empty islands by $q_i^\star(t)$, with $r_i(t)=1$ for the best island. The island-selection weights are
\begin{equation}
w_i^{\mathrm{isl}}(t)=1+\frac{|\mathcal{N}_t|-r_i(t)+1}{|\mathcal{N}_t|},
\qquad
p_i(t)=\frac{w_i^{\mathrm{isl}}(t)}{\sum_{j\in\mathcal{N}_t} w_j^{\mathrm{isl}}(t)}.
\end{equation}
Thus the best non-empty island has weight $2$, while the worst has weight $1+1/|\mathcal{N}_t|$, giving only a mild preference for higher-quality islands.

Two context entries are then sampled from the selected island. If that island contains fewer than two entries, sampling falls back to the full database. For a candidate context entry $e$ with quality $q_e$, the sampling weight is
\begin{equation}
w_e^{\mathrm{ctx}}=\exp\!\left(\frac{q_e-q_{\max}}{T_{\mathrm{ctx}}}\right),
\qquad
T_{\mathrm{ctx}}=20,
\end{equation}
where $q_{\max}$ is the maximum quality in the current sampling pool. Two entries are then drawn without replacement from the normalized probabilities
\begin{equation}
p(e)=\frac{w_e^{\mathrm{ctx}}}{\sum_{e'} w_{e'}^{\mathrm{ctx}}}.
\end{equation}

These entries are passed to the inner agent as reference fits. The returned proposal is evaluated and, if admitted, written back to the same database. The next prompt is therefore conditioned by previously accepted proposals from the ongoing search. This update rule defines the self evolving behavior. The recursive reuse of accepted proposals with increasing quality as future context gives the system a self-evolving context proven in self-reflection and self-refinement loops \cite{shinn2023reflexion,madaan2023selfrefine} and is also seen in prompt evolution and language-model-based optimization \cite{Promptbreeder,Yang0LLM}.

\subsection*{Metrics}

Not every proposal is retained; instead, admission explicitly balances solution quality and behavioral diversity, following the core principle of quality-diversity search \cite{MappingElites,Pugh2016QualityDA}. A candidate is admitted if its quality exceeds the 40th percentile of the selected island, if its diversity lies above the 80th percentile, or if no existing entry dominates it jointly in quality and diversity. Near duplicates are rejected in normalized parameter space.

Diversity is measured with a set of five metrics,
\begin{equation}
\mathbf{b} = \bigl[\chi^2_{\mathrm{img}},\;\hat{\sigma}_v,\;\theta_E,\;\gamma,\;F_{\mathrm{model}}\bigr]^{\!\mathsf{T}}\!,
\end{equation}
using the mean distance to nearby entries after column wise normalization. Islands are trimmed to a fixed size by removing the lowest quality entries. In this way, the database preserves diversity and provides the context for subsequent agent iterations.

\subsection*{Evaluation and physical verification}

Every candidate is scored by a composite objective,
\begin{equation}
\mathcal{Q} = -\alpha\,\mathrm{pen}(\chi^2_{\mathrm{img}}) - \delta\, R_{\mathrm{res}} - \beta\, \chi^2_{\mathrm{kin}} - \gamma\, P_{\mathrm{bnd}} - \epsilon\max(0,\,\mathrm{RMSE}_{\mathrm{Poisson}} - \tau).
\label{eq:quality}
\end{equation}
where $\mathrm{pen}(\chi^2_{\mathrm{img}})=|\chi^2_{\mathrm{img}}-1|$ penalizes deviation from the noise limited target, $R_{\mathrm{res}}$ measures residual autocorrelation, $\chi^2_{\mathrm{kin}} = (\hat{\sigma}_v - \sigma_v^{\mathrm{obs}})^2/(\delta\sigma_v^{\mathrm{obs}})^2$ compares predicted and observed velocity dispersion, $P_{\mathrm{bnd}}$ penalizes boundary solutions, and $\mathrm{RMSE}_{\mathrm{Poisson}}$ measures disagreement between $\kappa$ and $\tfrac{1}{2}\nabla^2\psi$.

This objective gives the system a rich scalar signal. The evaluator solves the imaging problem, predicts velocity dispersion through the kinematic module, measures residual structure, and checks numerical consistency of the potential and convergence. The kinematic term supplies the main physical anchor for the search. The Poisson term excludes numerically inconsistent mass models. Evaluations are executed under a hard timeout. If the kinematic solver stalls, the system falls back to imaging only scoring.

\subsection*{Agentic Deployment}

The same agent, proposal database and evaluator are reused across all fitting phases. Algorithm~\ref{alg:pipeline} summarizes the full procedure.

\begin{algorithm}
\caption{Self-evolving agent pipeline for lens fitting}\label{alg:pipeline}
\begin{algorithmic}[1]

\Require Observation $\mathcal{O}$, candidate model families $\mathcal{F}$, budget per pass $B_1, B_2, B_3$
\Ensure Best-fit parameters $\boldsymbol{\theta}^*$

\Statex \textbf{--- 1. AFMS: Model family search ---}
\State Run PSO scout on each $f \in \mathcal{F}$; rank by BIC; retain top $K$
\For{each retained family $f_k$}
    \State Initialize island database $\mathcal{D}_k$ with PSO seeds
\EndFor
\For{$t = 1, \ldots, B_1$}
    \State Select family $f_k$ via UCB: $S_k = \frac{1}{|\chi^2_{k,\mathrm{best}}-1|+\varepsilon} + c\sqrt{\frac{\log t}{n_k}}$
    \State \Call{EvolveFit}{$\mathcal{D}_k, \mathcal{O}$}
    \State Remove stagnating families from active pool
\EndFor
\State $f^* \gets$ family with best validated $\chi^2$; $\mathcal{D}^* \gets \mathcal{D}_{f^*}$

\Statex
\Statex \textbf{--- 2. PRL: Precision refinement (no bandit) ---}
\For{$t = 1, \ldots, B_2$} \Comment{same database $\mathcal{D}^*$ carried over}
    \State \Call{EvolveFit}{$\mathcal{D}^*, \mathcal{O}$} with tighter precision prompt
\EndFor

\Statex
\Statex \textbf{--- 3. RSI: Subhalo detection (no bandit) ---}
\State Compute pull map $P(\mathbf{x}) = (D - M)/\sigma$ from PRL best fit
\State Detect NFW subhalo candidates above threshold; fix PRL parameters.
\State Initialize subhalo database $\mathcal{D}_{\mathrm{sub}}$ seeded from PRL entries
\For{$t = 1, \ldots, B_3$}
    \State \Call{EvolveFit}{$\mathcal{D}_{\mathrm{sub}}, \mathcal{O}$}
\EndFor

\Statex
\Procedure{EvolveFit}{$\mathcal{D}, \mathcal{O}$} \Comment{shared across all passes}
    \State $j \gets$ \Call{PickIsland}{$\mathcal{D}$} \Comment{rank-weighted selection}
    \State $\mathbf{c}_1, \mathbf{c}_2 \gets$ \Call{SampleContext}{$\mathcal{D}, j$}
    \State $\boldsymbol{\theta}^+ \gets \varnothing$; \; $\chi^2_{\mathrm{best}} \gets \infty$
    \Statex \hspace{\algorithmicindent}\textit{--- Inner ReAct agent episode ---}
    \State Build prompt from $\mathcal{O}$ (image, PSF, noise), $\mathbf{c}_1, \mathbf{c}_2$, and $\sigma_v^{\mathrm{obs}}$
    \For{$s = 1, \ldots, T$}
        \State LLM generates reasoning trace and an \texttt{<action>} block
        \State Parse tool call: \textbf{evaluate} or \textbf{finish}
        \State Extract three candidate vectors $\{\boldsymbol{\theta}^{(1)}, \boldsymbol{\theta}^{(2)}, \boldsymbol{\theta}^{(3)}\}$
        \For{$m = 1, 2, 3$}
            \State Render forward model; compute $\chi^2_{\mathrm{img}}, \chi^2_{\mathrm{kin}}, R_{\mathrm{res}}, P_{\mathrm{bnd}}, \mathrm{RMSE}_{\mathrm{Poisson}}$
            \If{$\chi^2_{\mathrm{img}}(\boldsymbol{\theta}^{(m)}) < \chi^2_{\mathrm{best}}$}
                \State $\boldsymbol{\theta}^+ \gets \boldsymbol{\theta}^{(m)}$; \; $\chi^2_{\mathrm{best}} \gets \chi^2_{\mathrm{img}}(\boldsymbol{\theta}^{(m)})$
            \EndIf
        \EndFor
        \State Return scores, residual images, and visual analysis to LLM context
        \If{tool $=$ \textbf{finish}} \textbf{break} \EndIf
    \EndFor
    \Statex \hspace{\algorithmicindent}\textit{--- Admission into island database ---}
    \State $q \gets \mathcal{Q}(\boldsymbol{\theta}^+)$ \Comment{Eq.~\ref{eq:quality}}
    \State $\mathbf{b} \gets [\chi^2_{\mathrm{img}},\, \hat{\sigma}_v,\, \theta_E,\, \gamma,\, F_{\mathrm{model}}]$; \; $d \gets$ mean distance in normalized $\mathbf{b}$-space
    \If{$q > q_{40}(\mathcal{D}_j)$ \textbf{or} $d > d_{80}(\mathcal{D}_j)$ \textbf{or} non-dominated, \textbf{and not} duplicate}
        \State $\mathcal{D}_j \gets \mathcal{D}_j \cup \{\boldsymbol{\theta}^+\}$; \; trim island to max size
    \EndIf
\EndProcedure

\end{algorithmic}
\end{algorithm}

\paragraph{Model family search with bandit-based scheduling (AFMS).}
When several model families are plausible, a short PSO scout finds initial solutions and ranks them by BIC. This stage initializes the search over promising families. Agent budget is then allocated across the selected families by a UCB rule \cite{FINITE},
\begin{equation}
S_i = \frac{1}{|\chi^2_{i,\mathrm{best}}-1|+\varepsilon} + c\sqrt{\frac{\log N}{n_i}},
\end{equation}
where $n_i$ is the number of agent episodes assigned to family $i$ and $N$ is the total number of completed episodes. Each family maintains its own island database. Families that stagnate are removed from the active pool.

\paragraph{Precision Exploitation (PRL).}
After the competitive stage, the best validated family is refined further with the same database and a tighter numerical precision prompt. No reseeding occurs. The search continues from the accepted population already stored in that family.

\paragraph{Subhalo Detection from Residuals (RSI).}
The converged residuals are converted into a pull map $P(\mathbf{x}) = (D - M)/\sigma$. Significant peaks are used to augment the mass model with NFW subhalos. The same agent loop is then rerun on the observed system. Improvement is quantified by the reduced $\chi^2$, RMSE and the velocity dispersion value against the observed.

The same architecture therefore governs exploration, refinement and subhalo detection. The language model proposes parameters. The evaluator determines survival. The proposal database provides the evolving context for subsequent search. This coupling of reasoning, evaluation and memory update is the main methodological contribution of the pipeline.

\subsection*{Code Availability}
The code is available for review here at (\url{https://github.com/Leo-Fengxt/LensAgent})

\subsection*{Data Availability}
The HST images used in this project are FITS File  accessed from the MAST portal \url{https://mast.stsci.edu/}, Velocity dispersion data from Sloan Digital Sky Survey \url{https://www.sdss.org/science/}. The Lensing sample catalog is from the SLACS survey\cite{2008ApJ...682..964B}. 

\section*{Author Contribution}
X.F. led the research. The study was planned and directed by X.F., P.T. and Z.W..X.F., Z.W. and Z.S. conceived the project and developed the entire LensAgent pipeline. Z.W. developed the Kinematics Module. The data were analysed by Z.W. and Z.S., supported by X.F., J.-P.K. and P.T.. J.-P.K. provided astrophysical interpretation guidance. P.T. supervised the AI methodology. The manuscript was written by X.F., Z.W. and Z.S., after discussions with and 
input from all authors.

\subsection*{Acknowledgements} 
The authors would like to thank Junlin Han for fruitful discussions of the LLM agent framework and Shengyu He for helpful discussion of the physics idea of the project and the support throughout.
\subsection*{Competing Interests}
The authors declare no competing interests





\end{document}